\documentclass[12pt,psamsfonts]{amsart}
\usepackage{amsmath}
\usepackage{amsthm}
\usepackage{amssymb}
\usepackage{amscd}
\usepackage{amsfonts}
\usepackage{amsbsy}
\usepackage{epsfig}

\begin{document}

\title[Anomalous precession of the Mercury's
perihelion] {On the origin of the anomalous precession of
Mercury's perihelion}

\author[J. Gin\'{e}]
{Jaume Gin\'e}

\address{Departament de Matem\`atica, Universitat de Lleida,
Av. Jaume II, 69. 25001 Lleida, Spain}

\email{gine@eps.udl.es}

\thanks{The author is partially supported by a DGICYT
grant number BFM 2002-04236-C02-01 and by DURSI of Government of
Catalonia ``Distinci\'o de la Generalitat de Catalunya per a la
promoci\'o de la recerca universit\`aria".}


\date{}
\dedicatory{}

\maketitle

\begin{abstract}
Action at distance in Newtonian physics is replaced by finite
propagation speeds in classical post--Newtonian physics. As a
result, the differential equations of motion in Newtonian physics
are replaced by functional differential equations, where the delay
associated with the finite propagation speed is taken into
account. Newtonian equations of motion, with post--Newtonian
corrections, are often used to approximate the functional
differential equations. In \cite{G2} a simple atomic model based
on a functional differential equation which reproduces the
quantized Bohr atomic model was presented. The unique assumption
was that the electrodynamic interaction has a finite propagation
speed. In \cite{G3} a simple gravitational model based on a
functional differential equation which gives a gravitational
quantification and an explanation of the modified Titius--Bode law
is described. Firstly, in this work, we recall the calculations
made by Einstein to arrive at the explanation of the anomalous
precession of Mercury's perihelion. Secondly, we recover an
ancient work of Gerber in 1898 as a precursor of the retarded
theories, see \cite{Ge1}. In this paper Gerber gave an explanation
of the anomalous precession of the Mercury's perihelion in terms
of a velocity--dependent potential. In this paper an explanation
of the anomalous precession of Mercury's perihelion is given in
terms of a simple retarded potential, which, at first order,
coincides with Gerber's potential, and which agrees with the
author's previous works \cite{G2,G3}.
\end{abstract}

\section{Introduction}\label{s1}

The problem of the anomalous precession of the Mercury's
perihelion appeared in 1859 when the French astronomer Le Verrier
observed that the perihelion of the planet Mercury precesses at a
slightly faster rate than can be accounted the Newtonian mechanics
with the distribution of masses of the solar system well-known
until then. This discovery began different lines of investigation
to explain the new phenomenon. One of the explanations was the
existence of a new planet that would explain the anomaly in
Mercury's orbit within the context of Newton's laws. In others
lines of investigation it was considered the modification or
re-interpretation of the Newton's law of gravitation so that it
would give Mercury's precession with the known distribution of
masses of the solar system. For a complete description of the
historical development of the problem see \cite{M1,M2}.

Einstein found that the extra precession arises unavoidably from
the fundamental principles of General Relativity. The general
problem of the integration of the Einstein equations, given by
\[
R_{\mu \nu} - \frac{1}{2} R g_{\mu \nu} = \frac{8 \pi G}{c^4}
T_{\mu \nu},
\]
where $R_{\mu \nu}$ is the Ricci tensor, $R$ is the Ricci scalar,
$g_{\mu \nu}$ is the metric tensor and $T_{\mu \nu}$ is the
stress-energy tensor, is extremely difficult and the determination
of the explicit solutions is only possible in a restrict number of
cases. One of the most important is the Schwarzschild solution for
the case of a punctual mass or spherical and homogeneous and with
the assumption that the limit values in the infinite of the
$g_{\mu \nu}$ are the galilean values.

In \cite{M1} is given how to determine the relativistic prediction
for the advance of an elliptical orbit from the Schwarzschild
solution in a very comprehensible and clear form.

\section{Gerber's theory}

At the end of the 19th century, theoretical physicists were
investigating modifications of the Coulomb inverse--square law.
For instance, Gauss and Weber introduced velocity--dependent
potential to represent the electromagnetic field, consistent with
the finite propagation speed of changes in the field. The
application of this velocity--dependent potential to the
gravitation was immediately. Several physicists propose different
gravitational potentials based on finite propagation speed in
order to account the Mercury's orbital precession, see for
instance \cite{O1,O2} for a review of these proposals.

In fact, this line of investigation goes back to the works of
Laplace \cite{L} in 1805 where it is presented a correction of the
Newton force produced by the particle $m_1$ in $m_2$, which moves
with velocity $v$ given by
\[
{\bf F}= - G m_1 m_2  \left ( \frac{{\bf r}}{r^3} + \frac{{\bf
v}}{h} \right) ,
\]
where $h$ is the finite propagation speed. But this work didn't
find echo practically until the surroundings of 1880, when a
series of works to estimate the gravitational finite propagation
speed began. A brief list of authors that used the hypothesis of
the finite propagation speed is Th. von Oppolzer (1883), J. von
Hepperger (1889), R. Lehmann--Filhes (1894), K. Schwarzschild
(1900), H. Minkowski (1908), H. Poincar\'e (1908), W. Ritz (1909).
In other works different forms for the gravitational potential
were proposed; we can mention H. von Seeliger (1895) and C.
Neumann (1896). Under the influence of the electrodynamical
development made by F. Neumann (1845), W. Weber (1846) and B.
Riemann (1861), some authors began to think in modifying Newton's
law adding terms which depend on the speeds of the involved
bodies, see for instance \cite{W}. In 1870 F.G. Holzmuller
\cite{H} proposed a law of gravitation of the same form that the
electrodynamic Weber's law, given by
\[
F= \frac{Gm_1m_2}{r^2} \left ( 1 - \frac{\dot{r}^2}{h^2}  +
\frac{2 r \ddot{r}}{h^2}  \right).
\]
Later, F. Tisserand \cite{T} had used this law to study the
anomalous precession of Mercury's perihelion and he explained only
$14.1$ arc seconds per century. In the same way O. Liman (1886)
and M. L\'evi (1890), proposed a law of gravitation of the same
form that the electrodynamic Riemann's law, given by
\[
F=\frac{Gm_1m_2}{r^2} \left( 1 - \frac{
(\dot{r}_1-\dot{r}_2)^2}{h^2} \right) ,
\]
where $r_1$ and $r_2$ are the position vectors of the particles
$m_1$ and $m_2$, respectively. The Riemann--Liman--L\'evi law
explained only $28$ arc seconds per century of the anomalous
precession of Mercury's perihelion. Finally, M. L\'evi, by means
of a purely formal development, found a force law that led to the
observed exact value of the anomalous precession of Mercury's
perihelion. The theories to explain the form of the proposed law
forces are based, in general, in to do a parallelism between the
electromagnetism and the gravitation and to propose what is known
as gravitational field with a gravelectric component and with a
gravomagnetic component, see \cite{BN,Ma} and references therein.
Below, in the next section, we will see that all these laws are,
in fact, developments until certain order of a retarded potential.
These lines of research were abandoned when it was definitively
implanted Einstein's Relativity theory.

One of the first velocity--dependent potential used was
\[
V(r,\dot{r})=- \frac{m}{r} \frac{1}{ \left( 1- \frac{\dot{r}}{c}
\right)},
\]
where it is incorporate a finite propagation speed into the law of
gravity substituting the retarded radial distance for the present
distance. This velocity--dependent potential predicts only one
third of the observed value for the anomalous precession of
Mercury's perihelion, see \cite{M2}.

A German school teacher named Paul Gerber proposed in 1898 a
velocity--dependent potential that predicts exactly the observed
value for the anomalous precession of Mercury's perihelion, see
\cite{Ge1, Ge2}. In \cite{M2} it is concluded with a speculative
re--construction of a semi--classical line of reasoning by which
it is actually possible to derive Gerber's potential, albeit in a
way that evidently never occurred to Gerber. The proposed Gerber's
velocity--dependent potential is
\begin{equation}
V(r,\dot{r})=- \frac{m}{r} \frac{1}{ \left( 1- \frac{\dot{r}}{c}
\right)^2}. \label{eqn5}
\end{equation}
which depends not only on the radial distance from the
gravitational mass but also on the derivative (with respect to
time) of that distance. The force law associated to this
velocity--dependent potential is
\begin{eqnarray*}
f&=&\frac{d}{dt} \left( \frac{\partial V}{\partial \dot{r}}
\right)- \frac{\partial V}{\partial r} \\
&=& - \frac{m}{r^2} \left( 1- \frac{\dot{r}}{c} \right)^{-4}
\left( \frac{6r \ddot{r}}{c^2}-\frac{2 \dot{r}}{c}
\left(1-\frac{\dot{r}}{c} \right) + \left(1-\frac{\dot{r}}{c}
\right)^2 \right).
\end{eqnarray*}
and the expansion of this expression in powers of $\dot{r}/c$,
gives
\begin{equation}
f=-\frac{m}{r^2}\left(1-\frac{3\dot{r}^2}{c^2}+\frac{6r\ddot{r}}{c^2}
-\frac{8\dot{r}^3}{c^3}+\frac{24r\dot{r}\ddot{r}}{c^3}-\ldots
\right). \label{for}
\end{equation}

In \cite{M2}, it is showed that the Gerber's velocity--dependent
potential \eqref{eqn5} results in elliptical orbits that precess
by the same amounts as predicted by General Relativity (to the
lowest order of approximation), and of course these fact agrees
with the observed precession rates for the perihelia of the
planets, including Mercury. The question, then, is whether we can
justify the use of this particular velocity--dependent potential
rather than the Newtonian potential $V=-m/r(t)$.  Moreover, in
\cite{M2} it is also showed that although General Relativity and
Gerber's potential predict the same first-order precession, the
respective equations of motion are not identical, even at the
first non-Newtonian level of approximation. One of the objectives
in the Gerber's works, taking into account the assumption of a
finite propagation speed, was to infer the speed of gravity from
observations of the solar system. The open question is if gravity
and light move at the same speed, that it is still today on
discussing, see \cite{Wi} and references therein. In the
introduction of the Gerber's paper \cite{G2},
Ernst Gehrcke concludes:\\
\begin{quote}
Whether and how the theory of Gerber can be merged with the
well--known electromagnetic equations into a new unified theory is
a difficult problem, which still awaits a solution.
\end{quote}

\vspace{-0.25cm}

\section{A simple retarded potential}

Action at distance in Newtonian physics is replaced by finite
propagation speeds in classical post--Newtonian physics. As a
result, the differential equations of motion in Newtonian physics
are replaced by functional differential equations, where the delay
associated with the finite propagation speed is taken into
account. Newtonian equations of motion, with post--Newtonian
corrections, are often used to approximate the functional
differential equations, see, for instance,
\cite{Ch,Ch1,Ch2,Ch3,G1,R,R2}. In \cite{G2} a simple atomic model
based on a functional differential equation which reproduces the
quantized Bohr atomic model was presented. The unique assumption
was that the electrodynamic interaction has finite propagation
speed, which is a consequence of the Relativity theory. An
straightforward consequence of the theory developed in \cite{G2},
and taking into account that gravitational interaction has also a
finite propagation speed, is that the same model is applicable to
the gravitational 2-body problem. In \cite{G3} a simple
gravitational model based on a functional differential equation
which gives a gravitational quantification and an explanation of
the modified Titius--Bode law is described. In the following an
explanation of the anomalous precession of Mercury's perihelion is
given in terms of a simple retarded potential, which, at
first order, coincides with the Gerber's potential.\\

The most straightforward way of incorporating a finite propagation
speed into the law of gravity is to simplistically substitute the
present distance for the retarded radial distance, therefore, we
consider the simplest retarded potential
\begin{equation}
V=- \frac{m}{r(t-\tau)}, \label{eqs1}
\end{equation}
where $r(t)$ denotes the instantaneous position vector of the test
particle, at time $t$, and $\tau$ is the delay, so that
$r(t-\tau)$ is the retarded position of the test particle. In fact
this retarded potential depends on the position vector but also on
the velocity vector $\dot{r}$, on the acceleration vector
$\ddot{r}$ an so on. The appearance of a delay implies all these
dependences in the potential. From the retarded potential
\eqref{eqs1} we will obtain, in a theoretical point of view, the
equation of motion of the particle. This equation will be a
functional differential equation. The functional differential
equations of motion are generally difficult, often impossible, to
express in a form that is amenable to analysis. Thus, in order to
obtain useful dynamical predictions from realistic models, it is
frequent to replace the functional differential equations of
motion by approximations that are ordinary or partial differential
equations, see \cite{Ch}. In our case, if we develop the retarded
potential \eqref{eqs1} in powers of $\tau$ (up to second order in
$\tau $), we obtain
\begin{equation}
V \approx -\frac{m}{r} \left [ 1 + \frac{\dot{r}}{r} \ \tau +
\left( \frac{\dot{r}^2}{r^2}- \frac{\ddot{r}}{2r} \right) \ \tau^2
\right], \label{eqs2}
\end{equation}
To develop some easier calculations we can reject on the right
hand side of expression \eqref{eqs2} the term with $\ddot{r}$ (in
fact this term is negligible and only gives terms of higher
order). Hence, at this approximation, we obtain the
velocity--dependent potential
\begin{equation}
V \approx -\frac{m}{r} \left [ 1 + \frac{\dot{r}}{r} \ \tau +
\frac{\dot{r}^2}{r^2} \ \tau^2 \right], \label{eqs3}
\end{equation}
In a first approximation, the delay $\tau$ must be equal to $r/c$
(the time that the field uses to go from Mercury to the Sun at the
speed of the light) and according with the theories developed in
\cite{G2,G3}, we introduce a new constant $g$ in the delay and
hence, $\tau= g \, r/c$. Introducing this expression of the delay
in \eqref{eqs3} we have
\begin{equation}
V \approx -\frac{m}{r} \left [ 1 + g \frac{ r \dot{r}}{ c r} +
 g^2 \frac{ r^2 \dot{r}^2}{ c^2 r^2} \right]. \label{eqs3}
\end{equation}
On this basis, of this velocity-dependent potential function
\eqref{eqs3}, the gravitational force law is given by substituting
the potential function \eqref{eqs3} into equation
\begin{eqnarray*}
f&=&\frac{d}{dt} \left( \frac{\partial V}{\partial \dot{r}}
\right)- \frac{\partial V}{\partial r} =
-\frac{m}{r^2}\left(1-\frac{ g^2 \dot{r}^2}{c^2}+\frac{ 2 g^2
r\ddot{r}}{c^2} \right).
\end{eqnarray*}
It is easy to see that if we fix $g=\sqrt{3}$, we obtain the same
radial force, at first orders, that gives Gerber's potential, see
\eqref{for}. In fact, we have constructed a potential that,
varying $g$, predicts $2g^2 \pi m/(Lc^2)$ as non--Newtonian
advance of orbital perihelia per revolution, where $m$ is the
Sun's mass, $L$ is the semi--latus rectum of the orbit, and $c$ is
the speed of the light. Note that for $g=1$, it results in a value
which is one third of the observed value, so it predicts only
$14.1$ arc seconds per century for the precession of Mercury's
perihelion. The problem of the retarded potential \eqref{eqs1} is
that it can account for the anomalous precession of the Mercury's
perihelion precisely by adjusting a free parameter of the theory.
In the following we give a retarded potential which gives an
explanation of the anomalous precession of the Mercury's
perihelion without adjusting any free parameter of the theory. We
will see that this new retarded potential also coincides, at first
order, with Gerber's one.

We now consider a small modification of the retarded potential
\eqref{eqs1}, given by
\begin{equation}
V=- \frac{m}{r(t-\tau)} \frac{r(t)}{r(t-\tau)}, \label{eqs4}
\end{equation}
where the modification consists on dividing the retarded potential
\eqref{eqs1} by the quotient $r(t-\tau)/r(t)$. And this quotient
represents the ratio of the distance between the masses when the
potential was ``emitted" to distance between the masses at the
present instant. We can think that the retarded potential
\eqref{eqs1} was obtained from the Newtonian potential $V=-m/r(t)$
of the form
\[
V= -\frac{m} {r(t)\frac{r(t-\tau)}{r(t)}}=  -\frac{m} {r(t-\tau)},
\]
and the quotient $r(t-\tau)/r(t)$ is the corrective factor to
obtain the retarded distance. This corrective factor is applied
because the potential must propagate from the source to the
location particle in question. In the same way we can think that
the retarded potential \eqref{eqs4} is obtained from the Newtonian
potential $V=-m/r(t)$ of the form
\[
V= -\frac{m} {r(t)\frac{r(t-\tau)}{r(t)} \frac{r(t-\tau)}{r(t)}}=
-\frac{m} {r(t-\tau)} \frac{r(t)}{r(t-\tau)},
\]

In the same way that in the Neumann's theories \cite{N} we
conceive the potential essentially as information being
transmitted from place to place, and assumed a finite speed for
the propagation of this information.

\begin{figure}[htb]
\centerline{\hbox{
\epsfig{file=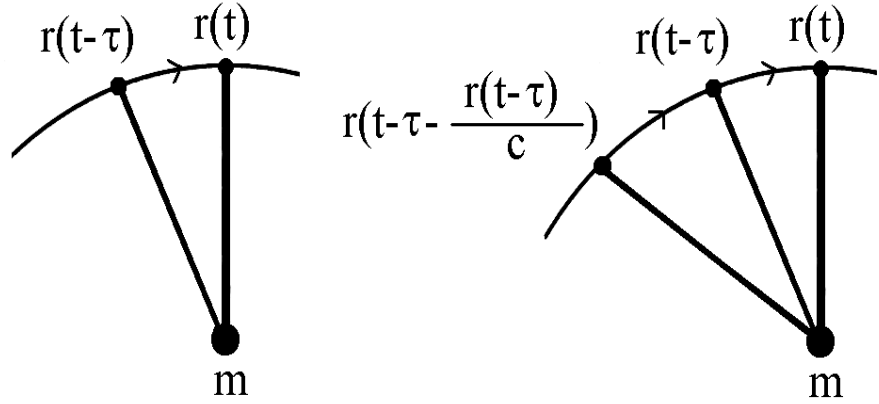,height=6.0cm,width=12.0cm} }} \caption{The
retarded position of the test particle.} \label{fig1}
\end{figure}
\par

As it is written in \cite{M2}, a particle sends forth a potential,
the value of which depends not only on the emitting particle, but
on the receiving particle. Therefore, the information must come
back from the receiving particle to the emitting particle. Thus we
ought to regard an elementary interaction not as a one-way
exchange, but as a two-way round-trip exchange. Hence, we must
apply the corrective factor twice in the initial potential.

In fact the correct expression of the retarded potential, taking
into account that the information must do a two-way round-trip and
that $\tau = r(t)/c$, is
\begin{equation}
V=- \frac{m}{r(t-\tau- \frac{r(t-\tau)}{c})}. \label{eqs10}
\end{equation}
where $r(t-\tau- r(t-\tau)/c)$ is the distance between the masses
when the potential was ``emitted" to go from the emitting particle
to the receiving particle and come back, see the second graphic of
Fig. 1. To find the retarded potential \eqref{eqs4} as
approximation of the retarded potential \eqref{eqs10} we take into
account that for a small $\tau$ we have that
\[
r(t)\, r(t-\tau- \frac{r(t-\tau)}{c}) \approx (r(t-\tau))^2 \,.
\]
Therefore, for a small $\tau$ we obtain
\[
V=- \frac{m}{r(t-\tau- \frac{r(t-\tau)}{c})} \approx -\frac{m}
{r(t-\tau)} \frac{r(t)}{r(t-\tau)}.
\]
Hence, the correct retarded potential is \eqref{eqs10}, but is a
functional potential which is difficult to express in a form that
is amenable to analysis. Therefore, we use the approximation
\eqref{eqs4} whose physical interpretation and use is totally
justified. In fact, the retarded potential \eqref{eqs4} is a
generalization of the Gerber's potential. The Gerber's potential
is the particular case when the velocity of the test particle is
constant, i.e., when $\ddot{r}=0$. In \cite{M2} a physical
explanation (albeit in a way that evidently never occurred to
Gerber) of the form of the Gerber's potential is given.

Now we are going to see that the retarded potential \eqref{eqs4}
gives an explanation of the anomalous precession of Mercury's
perihelion because coincides, at first order, with the force law
associated to Gerber's one. If we develop the retarded potential
\eqref{eqs4} in powers of $\tau$ (up to second order in $\tau $),
we obtain
\begin{equation}
V \approx -\frac{m}{r} \left [ 1 + \frac{ 2 \dot{r}}{r} \ \tau +
\left( \frac{3 \dot{r}^2}{r^2}- \frac{\ddot{r}}{r} \right) \
\tau^2 \right], \label{eqs5}
\end{equation}
To develop some easier calculations we can reject, as before, on
the right hand side of expression \eqref{eqs5} the term with
$\ddot{r}$ (in fact this term is negligible and only gives terms
of higher order). Hence, at this approximation, we obtain the
velocity--dependent potential
\begin{equation}
V \approx -\frac{m}{r} \left [ 1 + \frac{ 2 \dot{r}}{r} \ \tau +
\frac{3 \dot{r}^2}{r^2} \ \tau^2 \right], \label{eqs6}
\end{equation}
In a first approximation, the delay $\tau$ must be equal to $r/c$
(the time that the field uses to goes from Mercury to the Sun at
the speed of the light) according with the theories developed in
\cite{G2,G3}. Introducing this expression of the delay in
\eqref{eqs6} we have:
\begin{equation}
V \approx -\frac{m}{r} \left [ 1 + \frac{ 2 r \dot{r}}{ c r} +
 \frac{ 3 r^2 \dot{r}^2}{ c^2 r^2} \right]. \label{eqs7}
\end{equation}
On this basis, of this velocity-dependent potential function
\eqref{eqs7}, the gravitational force law is given by substituting
the potential function \eqref{eqs7} into the equation:
\begin{eqnarray*}
f&=&\frac{d}{dt} \left( \frac{\partial V}{\partial \dot{r}}
\right)- \frac{\partial V}{\partial r} =
-\frac{m}{r^2}\left(1-\frac{ 3 \dot{r}^2}{c^2}+\frac{ 6
r\ddot{r}}{c^2} \right).
\end{eqnarray*}
Hence, we obtain (without fixing any parameters) the same radial
force, at first orders, that gives Gerber's potential, see
\eqref{for}.

In fact, it is straightforward to see that, at first order, the
retarded potential \eqref{eqs4} and Gerber's potential coincide.
If we develop the retarded potential \eqref{eqs4} we have
\[
V=- \frac{m}{r(t-\tau)} \frac{r(t)}{r(t-\tau)}=-\frac{m}{r(t)-
\dot{r}(t) \tau + \ldots} \ \cdot \ \frac{r(t)}{r(t)-\dot{r}(t)
\tau + \ldots}
\]
\[
= \frac{m}{r(t)(1- \frac{\dot{r}(t)}{r(t)} \tau + \ldots)}  \
\cdot \ \frac{1}{1- \frac{\dot{r}(t)}{r(t)} \tau + \ldots} \, .
\]
Now substituting the delay $\tau= r/c$ we obtain
\[
V=-\frac{m}{r(t)(1- \frac{\dot{r}(t)}{c} + \ldots)}  \ \cdot \
\frac{1}{1- \frac{\dot{r}(t)}{c} + \ldots} \, .
\]
Therefore, at first order, the retarded potential \eqref{eqs4} has
the form
\[
V=-\frac{m}{r(t)\left ((1- \frac{\dot{r}(t)}{c})^2 + \ldots
\right)} \, .
\]
\section{Concluding remarks}

Therefore the anomalous precession of the Mercury's perihelion is,
in fact, to take into account the second order in the delay of the
retarded potential \eqref{eqs4} which is an approximation of the
correct retarded potential \eqref{eqs10}. It lacks to see if the
prediction for the deflection of electromagnetic waves grazing the
Sun using this potential coincide with value given by General
Relativity, assuming a plausible application of such potential to
the propagation of electromagnetic waves. We hope to give an
answer in a future work.\\

\noindent{\bf Acknowledgements:}

The author would like to thank Prof. M. Grau from Universitat de
Lleida for several useful conversations and remarks.

\end{document}